\documentclass[11pt]{article} %
\usepackage{nips14submit_e,times}
\usepackage{hyperref}
\usepackage{url}

\usepackage[left=1in, right=1in, top=0.99in, bottom=0.99in]{geometry} %

\usepackage[ruled, vlined]{algorithm2e}

\usepackage{amsfonts}

\usepackage{amsmath}

\usepackage{graphicx}

\usepackage{multicol}
\usepackage{multirow}
\usepackage{graphicx}
\usepackage{booktabs}

\usepackage{float}

\usepackage[style=numeric, backend=bibtex]{biblatex}
\newcommand{\km}{$k$-mer}

\newcommand{\phib}{\ensuremath{{\pmb \phi}}}

\newcommand{\Kcal}{\ensuremath{\mathcal{K}}}

\newcommand{\Rcal}{\ensuremath{\mathcal{R}}}
\newcommand{\Scal}{\ensuremath{\mathcal{S}}}

\newcommand{\toolname}{Kover}

\newcommand{\mtub}{\textit{My\-co\-bac\-te\-rium tu\-ber\-cu\-lo\-sis}}

\newcommand{\pas}{\textit{P.\ ae\-ru\-gi\-no\-sa}}

\newcommand{\mtubs}{\textit{M.\ tu\-ber\-cu\-lo\-sis}}
\newcommand{\spneus}{\textit{S.\ pneu\-mo\-ni\-ae}}

\newcommand{\abs}{\textit{A.\ bau\-man\-nii}}

\newcommand{\eqdef}{\overset{{\rm \mbox{\tiny def}}}{=}}

\makeatletter
\def\@xfootnote[#1]{%
  \protected@xdef\@thefnmark{#1}%
  \@footnotemark\@footnotetext}
\makeatother

\newcommand{\drouin}{Drouin}

\title{Large scale modeling of antimicrobial resistance\\with interpretable classifiers}

\author{}

\nipsfinalcopy %

\begin{document}
	
	\maketitle
	
	{\vspace{-22mm} \large
    \begin{center}
    Alexandre Drouin$^{1,3,}$\footnote[$\dagger$]{Corresponding author: \href{mailto:alexandre.drouin.8@ulaval.ca}{alexandre.drouin.8@ulaval.ca}\\{\indent Peer-reviewed and accepted for presentation at the Machine Learning for Health Workshop, NIPS 2016, Barcelona, Spain.}}, Fr\'ed\'eric Raymond$^{2,3}$, Ga\"el Letarte St-Pierre$^{1,3}$, Mario Marchand$^{1,3}$,\\ Jacques Corbeil$^{2,3}$, Fran{\c c}ois Laviolette$^{1,3}$
    \end{center}}
    
    \begin{center}\small \vspace{-2mm}
    ${}^1$ Department of Computer Science and Software Engineering, ${}^2$  Infectious Disease Research Center, ${}^3$ Big Data Research Center\\Universit\'e Laval, Qu\'{e}bec, Canada
    \vspace{2.5mm}
    \end{center}
	
	\begin{abstract}
		Antimicrobial resistance is an important public health concern that has implications in the practice of medicine worldwide.
		Accurately predicting resistance phenotypes from genome sequences shows great promise in promoting better use of antimicrobial agents, by determining which antibiotics are likely to be effective in specific clinical cases.
		In healthcare, this would allow for the design of treatment plans tailored for specific individuals, likely resulting in better clinical outcomes for patients with bacterial infections.
		In this work, we present the recent work of \drouin~et al. (2016) on using Set Covering Machines to learn highly interpretable models of antibiotic resistance and complement it by providing a large scale application of their method to the entire PATRIC database.
		We report prediction results for $36$ new datasets and present the Kover AMR platform, a new web-based tool allowing the visualization and interpretation of the generated models.
	\end{abstract}

	\section{Introduction}
	
	Modern medicine relies on antimicrobial drugs to treat infections. 
	However, bacteria have evolved mechanisms to protect themselves from antibiotics~\cite{walsh2000, munita2015}. 
	Thus, the use and abuse of antibiotics has lead to the selection and to the spread of antibiotic-resistant pathogens. 
	In order to define the right treatment and to reduce clinical failures, the prediction of antimicrobial resistance (AMR) is essential in choosing the right drugs to treat a specific patient~\cite{friedman2016}. 
	In clinical laboratories, AMR is measured using antibiograms. 
	This method determines the minimal inhibitory concentration (MIC) of an antibiotic by measuring the growth of the infecting microorganism in presence of different concentrations of the drug.
	The overall objective of this method is to determine if the pathogen will respond to treatment. 
	Clinicians do so by comparing the measured MIC to the guidelines from CLSI or EUCAST, which are constantly being reevaluated by international committees~\cite{kahlmeter2015}.
	However, while susceptibility to antibiotics can be predicted using MIC, it does not always hold true, as susceptible isolates can sometimes become phenotypically resistant given the proper conditions~\cite{martineau2000}. 
	In this case, genomic determinants of resistance can be effective in predicting clinically relevant antimicrobial resistance~\cite{mcarthur2015}. 
	In addition, determination of the MIC requires the growth of microorganisms, which generally necessitates one to two days of \textit{in vitro} culture, and even more in the case of slow-growing organisms, such as \mtub~\cite{witney2016}. 
	Genomic methods, such as polymerase chain reaction or whole genome sequencing, can now be used to predict the resistance phenotypes of pathogens in a more rapid manner~\cite{punina2015}. 
	
	Reanalysis of publicly available genome databases for which AMR phenotypes are available is a useful starting point to improve our understanding of the relationship between genotype and phenotype. Indeed, several groups have used machine learning and statistics to understand and predict antimicrobial resistance (\cite{bradley2015, earle2016, santerre2016, drouin2016, davis2016}). However, models that predict AMR should not be static and should improve as new genomes are added to databases. For instance, the Pathosystems Resource Integration Center (PATRIC) database is a large-scale aggregation platform for bacterial genomes and their associated metadata~\cite{wattam2013, santerre2016}. The number of genomes in the  PATRIC database nearly doubled between 2014 and 2015, with now more than 52 thousand microbial genome sequences. Recently, \drouin~et al. (2016) proposed to use the Set Covering Machine~\cite{marchand2002} to learn extremely sparse models of antimicrobial resistance that are intelligible for domain experts \cite{drouin2016}. They compared their models to more complex predictors, such as linear and kernel-based Support Vector Machines~\cite{cortes1995, scholkopf2004, shawetaylor2004}, as well as decision trees~\cite{breiman1984}, and showed that Set Covering Machines achieved comparable, often superior, generalization performance, while being significantly sparser.
	Moreover, they demonstrated that highly accurate models of AMR could be obtained, despite features spaces tens of thousands of times larger than the number of learning examples.
	
	The present work summarizes the work of \drouin~et al. (2016) and presents extensive new results. Specifically, we present a large scale application of their method to prospectively generate predictive models of AMR from the ever-growing collection of genomes in the PATRIC database.
	Moreover, we present the Kover AMR Platform (\href{https://aldro61.github.io/kover-amr-platform/}{https://aldro61.github.io/kover-amr-platform/}), a web-based tool that catalogs AMR prediction results for a wide variety of species and antibiotics, providing detailed metrics and allowing the visualization of the generated models.
	This initiative will allow the interpretation of our results by healthcare researchers, generating new research and treatment opportunities.

	\section{Methods}
	
	\subsection{Problem statement}
	
	We address the problem of predicting antimicrobial resistance as a supervised learning problem.
	The goal is to learn a model that accurately discriminates genomes that are resistant or susceptible to an antibiotic based on genomic characteristics.
	Formally, we assume that we are given a dataset $S \eqdef \{(x_i, y_i)\}_{i=1}^m \sim D^m$, where $x_i \in \{A, C, G, T\}^*$ is a bacterial genome, $y_i \in \{0, 1\}$ is its associated phenotype ($0$ for susceptible and $1$ for resistant) and $D$ is an unknown data generating distribution from which the dataset is sampled.
	We start by defining an alternative representation for the genomic sequences, where each genome is characterized by the presence or absence of every possible \km, i.e. every possible sequence of $k$ DNA nucleotides. 
	This representation is obtained through a mapping function $\phib: \{A, C, G, T\}^* \rightarrow \{0, 1\}^{4^k}$, such that $\phi_j(x) \eqdef 1$ if the \km~$k_j$ is in the genome $x$ and $0$ otherwise.
	This yield the transformed dataset $S' \eqdef \{(\phib(x_i), y_i)\}_{i=1}^m$, which is then used to train the learning algorithm.
	
	The goal is then to find a model $h$ that has a good generalization performance, i.e. that minimizes the probability $R(h)$ of making a prediction error for any example drawn according to distribution $D$, i.e., 
	\begin{equation}\label{eq:true_risk}
	R(h) \eqdef \underset{(x, y) \sim D}{Pr} [h(\phib(x)) \not= y].
	\end{equation}
	Furthermore, we seek highly interpretable models from which biologically relevant knowledge can be extracted.

	\subsection{The Set Covering Machine}
	
	Such interpretable models are obtained through the Set Covering Machine algorithm (SCM)~\cite{marchand2002, drouin2016}, which produces models that are logical combinations (conjunctions or disjunctions) of boolean-valued rules that are generated from the data. We now briefly present the algorithm and direct the reader to \cite{drouin2016, marchand2002} for further explanations.
	
	The input of the SCM algorithm is a set of learning examples $\Scal$ and a set of boolean-valued rules $\Rcal$.
	In our context, $\Scal$ is composed of genomes, in the \km~form induced by $\phib$, and their labels.
	Let $\Kcal$ be the set of all, possibly overlapping, \km s that are present in at least one genome of $\Scal$.
	For each \km~$k_j \in \Kcal$, we consider a presence rule, defined as $p_{k_j}(\phib(x)) \eqdef I[\phi_j(x) = 1]$ and an absence rule, defined as $a_{k_j}(\phib(x)) \eqdef I[\phi_j(x) = 0]$, where $I[true] = 1$ and $0$ otherwise.
	These boolean-valued rules constitute the set $\Rcal$.
	Given $\Scal$ and $\Rcal$, the SCM attempts to find the model that relies on the smallest possible set of rules $\Rcal^\star = \{r^\star_1, ..., r^\star_n\} \subseteq \Rcal$, while minimizing Equation (\ref{eq:true_risk}).
	The models generated can be conjunctions $h(\phib(x)) \eqdef r^\star_1(\phib(x)) \wedge ... \wedge r^\star_n(\phib(x))$ or disjunctions $h(\phib(x)) \eqdef r^\star_1(\phib(x)) \vee ... \vee r^\star_n(\phib(x))$. Hence, they directly highlight the importance of a small set of genomic sequences for predicting AMR phenotypes.
	
	However, it is important to note that the distribution $D$ is unknown; therefore, it is not possible to directly minimize Equation (\ref{eq:true_risk}). 
	Instead, the algorithm constructs a model that achieves an appropriate trade-off between the empirical risk (i.e., the fraction of training errors) and the number of rules it uses. A model containing many rules is likely to overfit the data, whereas a model containing too few rules is likely to underfit.   
	To find the appropriate trade-off between the classifier's size and its accuracy on the training set, the SCM relies on a modified version of the set covering greedy algorithm of Chv\'atal~\cite{chvatal1979}, which has a worst-case guarantee. 
	The running time and space complexities of this algorithm are linear in the number of examples and rules, thus linear in the number of genomes and \km s. 
	Consequently, this algorithm is particularly well-suited for learning from large datasets of extreme dimensionalities, such as the ones that often occur in healthcare applications.

	The experiments in this work were performed using \toolname, the SCM implementation of \drouin~et al. (2016), which has been tailored for learning from \km~represented genomes~\cite{drouin2016}.
	In \toolname, the SCM algorithm is trained \textit{out-of-core}, which means that the data resides on the disk and is accessed in blocks by the algorithm.
	The implementation exploits a compressed data representation and atomic CPU instructions to speed up computations.
	It is open-source and available at \href{https://github.com/aldro61/kover}{https://github.com/aldro61/kover}.
	
	\subsection{Data acquisition}
	
	The data used in this study were extracted from the PATRIC database via its FTP server (\href{ftp://ftp.patricbrc.org}{ftp.patricbrc.org}).
	First, the latest AMR metadata were acquired.
	These data consisted of genome identifiers and measured resistance phenotypes (resistant or susceptible) for various antibiotics.
	The data were segmented by species and antibiotic to form datasets, and those with at least $50$ genomes in each class were retained.
	Finally, for each dataset, the assembled genomes were downloaded and their \km~representation was obtained using the DSK \km~counter~\cite{rizk2013}.
	The value of $k$ was set to $31$, since extensive experimentation has shown that this value is appropriate for the task at hand~\cite{drouin2016}.

	\section{Results and discussion}
	
	\begin{figure}
		\centering
		\includegraphics[width=\textwidth]{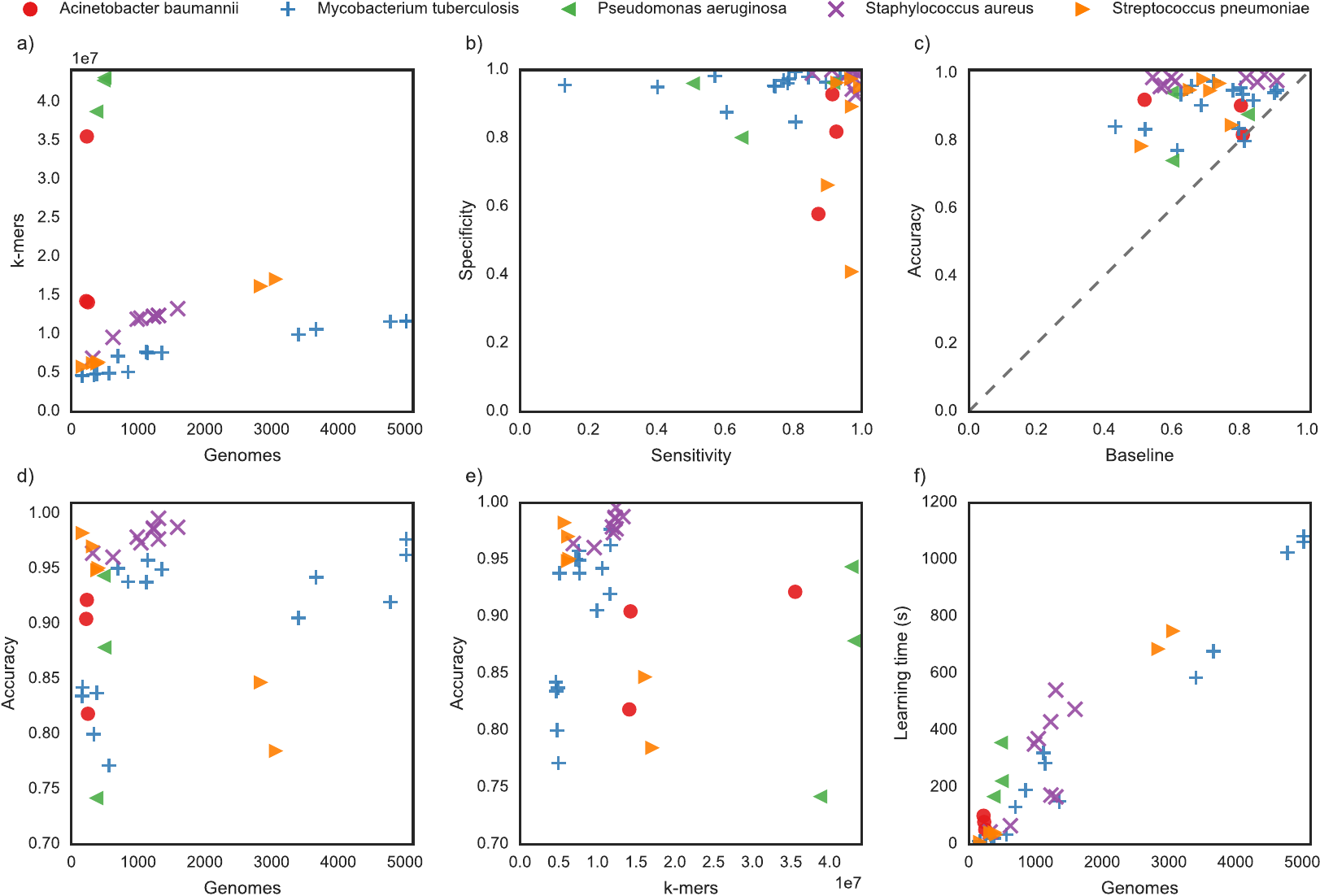} \vspace{-5mm}
		\caption{Each point corresponds to a dataset. Colored symbols identify bacterial species. a) Size of the datasets in terms of genomes and \km s; b) Specificity of the models with respect to their sensitivity; c) Accuracy of the models with respect to the baseline predictor (see text for details); d) Accuracy of the models with respect to the number of genomes; e) Accuracy of the models with respect to the number of \km s; f) Time required to train the algorithm with respect to the number of genomes. \vspace{-0.2cm}}
		\label{fig:results}
	\end{figure}
	
	A total of $36$ datasets, comprising five bacterial species, were extracted from PATRIC.
	A detailed list is available in Appendix - Table~1.
	Figure~\ref{fig:results}a shows the size of these datasets, in which the number of examples ranged from $161$ to $5022$ and the number of \km s from $4.6$ to $43.1$ millions.
	For most species, an increase in the number of genomes does not lead to a large increase in the number of \km s.
	This reflects the limited genomic diversity that exists within species (e.g.: \mtubs~\cite{hershberg2008}).
	Contrasting results were observed for \pas~and \abs, which have higher genomic diversities~\cite{kos2015, wallace2016}.

	The SCM's ability to predict AMR phenotypes was evaluated on held-out subsets of the data.
	First, each dataset was randomly partitioned into a training set $\Scal$ ($80\%$) and a testing set $\mathcal{T}$ ($20\%$).
	The SCM was then trained on $\Scal$ and metrics were computed based on the model's predictions on $\mathcal{T}$.
	This procedure was repeated $10$ times, on different partitions, and the resulting metrics were averaged.
	The values of the algorithm's hyperparameters\footnote{\scriptsize $p \in \{0.1, 0.178, 0.316, 0.562, 1, 1.778, 3.162, 5.623, 10, +\infty \}, s \in \{1,...,10\}$, model\_type $\in \{\mbox{conjunction}, \mbox{disjunction}\}$ (notation of \cite{drouin2016})} (HPs) were selected by bound selection (see~\cite{marchand2002, drouin2016}).
	Bound selection uses a probabilistic upper bound on Equation (\ref{eq:true_risk}), computed from the training data, to score each of the HP combinations.
	For each of the latter, a single training of the algorithm is required; hence, bound selection is much faster than standard cross-validation.
	\drouin~et al (2016). proposed such a bound for conjunctions/disjunctions of presence/absence rules of \km s and found that bound selection yielded results comparable to those of 5-fold cross-validation.
	
	The results are summarized in Figure~\ref{fig:results}b-e and detailed in Appendix - Table 2.
	Figure~\ref{fig:results}b shows the specificity of the models with respect to their sensitivity.
	A perfect model would score $1$ for each of these metrics.
	Specificities superior to $80\%$ were achieved for $33/36$ datasets and comparable sensitivities were achieved for $25/36$ datasets.
	In general, the obtained models are more specific than sensitive, meaning that they sometimes fail to identify resistant isolates, but very rarely mark a susceptible isolate as resistant.
	Figure~\ref{fig:results}c compares the SCM models to a baseline predictor that predicts the most abundant class in the training data.
	The SCM models, which achieve accuracies greater than $80\%$ on $33/36$ datasets, generally surpass the baseline predictors, indicating that the algorithm extracts relevant patterns of antibiotic resistance.
	Of note, the models learned by the SCM are extremely sparse, using an average of $2.5$ rules (std: $2.2$), which makes them well-suited for further review and experimental validation.
	Moreover, Figure~\ref{fig:results}c highlights the strong class imbalance that exists in some of the datasets considered, as the baseline predictors often achieve high accuracies.
	Furthermore, based on Figure~\ref{fig:results}d, we observe that the accuracy of the models is generally higher for datasets that contain more examples.
	There are notable exceptions, such as \spneus, where the accuracies are lower for larger datasets.
	However, the two largest datasets for this species correspond to combinations of drugs (Beta-lactams and Trimethoprim/Sulfamethoxazole), which could complexify the learning task.
	Also, notice that there are cases where the algorithm achieves near perfect accuracies with very few examples, despite disproportionately large feature spaces.
	In fact, Figure~\ref{fig:results}e illustrates that the accuracy of the models is not related to the number of \km s.
	This supports the theoretical and empirical results of \drouin~et al. (2016), which found that the SCM could avoid overfitting, even in such high dimensional settings.
	Finally, Figure~\ref{fig:results}f shows the time required to train the algorithm with respect to the number of genomes in each dataset.
	The time, which grows linearly with the number of genomes, varied between $8$ seconds and $18$ minutes, using a single CPU core and less than 1 GB of memory.

	The models generated using the SCM are available through the Kover AMR Platform.
	The \km~sequences of the rules in these models were annotated using BLAST~\cite{altschul1990}. 
	This revealed that, for most antibiotics, the SCM was accurate in identifying known resistance mechanisms. For example, the absence of one \km~located in the DNA gyrase subunit A (\textit{gyrA}) predicts resistance to moxifloxacin in \mtubs~with an error rate as low as $4\%$. This \km~refers to the amino-acids 88 to 94 of GyrA, the mutation of which confers resistance to fluoroquinolones~\cite{chien2016}. Thus, the model relies on the absence of the susceptible genotype to account for the presence of resistant variants, which are more diverse. 
	This behavior was also observed for predicting resistance to isoniazid in \mtubs, where the model rightly targets a region of the \textit{katG} gene where multiple mutations are known to induce resistance~\cite{cade2010isoniazid, da2011molecular}.
	The model also relies on a \km~in the \textit{rpoB} gene, a known rifampicin resistance determinant~\cite{da2011molecular}.
	This could result from the frequent combined use of antituberculosis drugs.
	
	We have briefly demonstrated the accuracy and interpretability of models produced by the SCM.
	Detailed results are available in the Appendix and the Kover AMR platform, which allows the visualization and further investigation of AMR models generated at an unprecedented scale.

	\section{Conclusion}
	
	In summary, we have outlined the recently published work of \drouin~et al. (2016), while complementing their analysis with a large-scale application of their method to the ever-growing PATRIC database.
	Kover, an extremely efficient out-of-core implementation of the SCM, allowed the rapid generation of these results with limited computational resources.
	Moreover, our results show that the method yields accurate results for predicting AMR in most datasets and that, due to their strong interpretability, the obtained models can generate biologically relevant insight into these phenotypes.
	
	On another hand, contrasting results were obtained, which pave the way to extensions of their method.
	In fact, the obtained models were generally highly specific, but some lacked sensitivity.
	This could result from seeking the sparsest model that detects resistant genomes in the entire population of isolates.
	Hence, the deconvolution of resistance mechanisms based on population structure could provide a deeper understanding of antibiotic resistance and increase the sensitivity of the models.
	Future work will therefore involve the development of algorithmic extensions to the Set Covering Machine, which allow the inclusion of prior knowledge of the population structure and the biological structures present in the data (e.g., gene functions, pathways).
	
	We have only scratched the surface of the biological knowledge that can be generated from these results.
	The fact that our approach generates interpretable predictors, together with our proposed Kover AMR Platform (\href{https://aldro61.github.io/kover-amr-platform/}{https://aldro61.github.io/kover-amr-platform/}) will allow further analysis of these results by researchers with diverse backgrounds, bridging the gap between machine learning and healthcare research.
	
	\section*{Acknowledgements}
	The authors acknowledge S\'ebastien Gigu\`ere and Pier-Luc Plante for helpful comments and suggestions.
    Computations were performed on the Colosse supercomputer at Universit\'{e} Laval (resource allocation project: nne-790-ae), under the auspices of Calcul Qu\'{e}bec and Compute Canada. 
    AD is recipient of an Alexander Graham Bell Canada Graduate Scholarship Doctoral Award of the National Sciences and Engineering Research Council of Canada (NSERC). 
    This work was supported in part by the NSERC Discovery Grants (FL; 262067, MM; 122405).
    JC acknowledges the Canada Research Chair in Medical Genomics.
	
	\printbibliography
\end{document}